\documentclass[10pt,conference,nofonttune]{IEEEtran}

\usepackage[acronyms,nonumberlist,nopostdot,nomain,nogroupskip]{glossaries}
\usepackage[numbers,sort&compress]{natbib}
\usepackage{xspace}

\AtBeginDocument{%
	\providecommand\BibTeX{{%
			\normalfont B\kern-0.5em{\scshape i\kern-0.25em b}\kern-0.8em\TeX}}}

\usepackage[utf8]{inputenc}
\usepackage{xcolor}
\usepackage{amsmath}
\usepackage{amssymb}
\usepackage{bm}

\usepackage[acronyms,nonumberlist,nopostdot,nomain,nogroupskip]{glossaries}
\usepackage{booktabs}
\usepackage{lipsum}
\usepackage{tabularx}
\usepackage[noend]{algorithmic}
\usepackage{algorithm, setspace}
\usepackage{tikz}
\usepackage{pgfplots}
\pgfplotsset{compat=newest}
\pgfplotsset{plot coordinates/math parser=false}
\newlength\fheight
\newlength\fwidth
\usetikzlibrary{plotmarks,patterns,decorations.pathreplacing,backgrounds,calc,arrows,arrows.meta,spy,matrix}
\usepgfplotslibrary{patchplots,groupplots,colorbrewer}
\usepackage{tikzscale}
\usepackage{colortbl}
\usepackage{array}
\usepackage[draft]{hyperref}
\usepackage{balance}
\usepackage{soul}

\newif\ifexttikz
\exttikzfalse

\ifexttikz
\usetikzlibrary{external}
\fi

\usepackage{multirow}

\renewcommand{\figurename}{Fig.}
\usepackage[font=footnotesize]{subcaption}
\usepackage[font=footnotesize]{caption}

\usepackage{mathtools}

\newcommand{\beq}{\begin{equation}}
	\newcommand{\eeq}{\end{equation}}
\newcommand{\bit}{\begin{itemize}}
	\newcommand{\eit}{\end{itemize}}

\newcommand{\esword}{\textit{eSWORD}\xspace}

\tikzstyle{startstop} = [rectangle, rounded corners, minimum width=2cm, minimum height=0.5cm,text centered, draw=black]
\tikzstyle{io} = [trapezium, trapezium left angle=70, trapezium right angle=110, minimum width=3cm, minimum height=1cm, text centered, draw=black]
\tikzstyle{process} = [rectangle, minimum width=2cm, minimum height=0.5cm, text centered, draw=black, alignb=center]
\tikzstyle{decision} = [ellipse, minimum width=2cm, minimum height=1cm, text centered, draw=black]
\tikzstyle{arrow} = [thick,<->,>=stealth]
\tikzstyle{line} = [thick,>=stealth]
\tikzstyle{darrow} = [thick,<->,>=stealth,dashed]
\tikzstyle{sarrow} = [thick,->,>=stealth]
\tikzstyle{larrow} = [line width=0.1mm,dashdotted,->,>=stealth]

\makeatletter
\def\grd@save@target#1{%
	\def\grd@target{#1}}
\def\grd@save@start#1{%
	\def\grd@start{#1}}
\tikzset{
	grid with coordinates/.style={
		to path={%
			\pgfextra{%
				\edef\grd@@target{(\tikztotarget)}%
				\tikz@scan@one@point\grd@save@target\grd@@target\relax
				\edef\grd@@start{(\tikztostart)}%
				\tikz@scan@one@point\grd@save@start\grd@@start\relax
				\draw[minor help lines] (\tikztostart) grid (\tikztotarget);
				\draw[major help lines] (\tikztostart) grid (\tikztotarget);
				\grd@start
				\pgfmathsetmacro{\grd@xa}{\the\pgf@x/1cm}
				\pgfmathsetmacro{\grd@ya}{\the\pgf@y/1cm}
				\grd@target
				\pgfmathsetmacro{\grd@xb}{\the\pgf@x/1cm}
				\pgfmathsetmacro{\grd@yb}{\the\pgf@y/1cm}
				\pgfmathsetmacro{\grd@xc}{\grd@xa + \pgfkeysvalueof{/tikz/grid with coordinates/major step x}}
				\pgfmathsetmacro{\grd@yc}{\grd@ya + \pgfkeysvalueof{/tikz/grid with coordinates/major step y}}
				\foreach \x in {\grd@xa,\grd@xc,...,\grd@xb}
				\node[anchor=north] at (\x,\grd@ya) {\pgfmathprintnumber{\x}};
				\foreach \y in {\grd@ya,\grd@yc,...,\grd@yb}
				\node[anchor=east] at (\grd@xa,\y) {\pgfmathprintnumber{\y}};
			}
		}
	},
	minor help lines/.style={
		help lines,
		gray,
		line cap =round,
		xstep=\pgfkeysvalueof{/tikz/grid with coordinates/minor step x},
		ystep=\pgfkeysvalueof{/tikz/grid with coordinates/minor step y}
	},
	major help lines/.style={
		help lines,
		line cap =round,
		line width=\pgfkeysvalueof{/tikz/grid with coordinates/major line width},
		xstep=\pgfkeysvalueof{/tikz/grid with coordinates/major step x},
		ystep=\pgfkeysvalueof{/tikz/grid with coordinates/major step y}
	},
	grid with coordinates/.cd,
	minor step x/.initial=.5,
	minor step y/.initial=.2,
	major step x/.initial=1,
	major step y/.initial=1,
	major line width/.initial=1pt,
}
\makeatother

\newacronym{6g}{6G}{sixth generation}
\newacronym{5g}{5G}{fifth generation}
\newacronym{snr}{SNR}{signal-to-noise ratio}
\newacronym{sinr}{SINR}{signal-to-interference-plus-noise ratio}
\newacronym{bw}{BW}{bandwidth}
\newacronym{mod}{Mod}{Modulation}
\newacronym[plural=\gls{cnn}s,firstplural=convolutional neural networks (CNNs)]{cnn}{CNN}{convolutional neural network}

\newacronym{iq}{I/Q}{in phase/quadrature}
\newacronym{mbc}{MBC}{modulation and bandwidth classification}
\newacronym{ml}{ML}{machine learning}
\newacronym{phy}{PHY}{physical layer}
\newacronym{cvl}{CVL}{convolutional layer}
\newacronym[plural=\gls{dnn}s,firstplural=deep neural networks (DNNs)]{dnn}{DNN}{deep neural network}
\newacronym{mmwave}{mmWave}{millimeter wave}
\newacronym{dsp}{DSP}{digital signal processing}
\newacronym{dsa}{DSA}{dynamic spectrum access}
\newacronym{ism}{ISM}{industrial, scientific and medical}
\newacronym{csi}{CSI}{channel state information}
\newacronym{fcc}{FCC}{Federal Communication Commission}
\newacronym{rfp}{RFP}{radio fingerprinting}
\newacronym{sdr}{SDR}{software-defined radio}

\newacronym{pus}{PUs}{primary users}
\newacronym{sus}{SUs}{secondary users}
\newacronym{iot}{IoT}{Internet of Things}
\newacronym{mimo}{MIMO}{multi-input, multi-output}
\newacronym{mum}{MU-MIMO}{multi-user \gls{mimo}}
\newacronym{sum}{SU-MIMO}{single-user \gls{mimo}}
\newacronym{iui}{IUI}{inter-user interference}
\newacronym{isi}{ISI}{inter-stream interference}

\newacronym{wlan}{WLAN}{Wireless LAN}
\newacronym{wlans}{WLANs}{Wireless Local Area Networks}
\newacronym{rlnc}{RLNC}{Random Linear Network Coding}
\newacronym{drx}{DRX}{Discontinuous Reception}
\newacronym{cpu}{CPU}{Central Processing Unit}
\newacronym{soc}{SoC}{system-on-chip}
\newacronym{dcm}{DCM}{distributed cooperative \gls{mimo}}
\newacronym{comp}{CoMP}{Coordinated Multi-Point}
\newacronym{ap}{AP}{access point}
\newacronym{sta}{STA}{station}
\newacronym{dl}{DL}{downlink}

\newacronym{fir}{FIR}{Finite Impulse Response}

\newacronym{mcs}{MCS}{modulation and coding scheme}
\newacronym{cfr}{CFR}{channel frequency response}
\newacronym{ndp}{NDP}{null data packet}
\newacronym[plural=\gls{ltf}s,firstplural=long training fields (LTFs)]{ltf}{LTF}{long training field}
\newacronym{vht}{VHT}{very high throughput}
\newacronym{ofdm}{OFDM}{orthogonal frequency-division multiplexing}
\newacronym{cfo}{CFO}{carrier frequency offset}
\newacronym{sfo}{SFO}{sampling frequency offset}
\newacronym{pdd}{PDD}{packet detection delay}
\newacronym{ppo}{PPO}{phase-locked loop offset}
\newacronym{pa}{PA}{phase ambiguity}
\newacronym{sbc}{SBC}{single board computer}

\newacronym[plural=\gls{cm}s,firstplural=confusion matrices (CMs)]{cm}{CM}{confusion matrix}

\newacronym{id}{ID}{identifier}
\newacronym{aoa}{AoA}{angle of arrival}
\newacronym{ul}{UL}{uplink}

\newacronym{svd}{SVD}{singular value decomposition}

\newacronym[plural=\gls{pdf}s,firstplural=probability density functions (PDFs)]{pdf}{PDF}{probability density function}

\newacronym{mse}{MSE}{mean-square-error}
\newacronym{mmse}{MMSE}{minimum \gls{mse}}
\newacronym{kkt}{KKT}{Karush-Kuhn-Tucker}
\newacronym{fhss}{FHSS}{frequency hopping spread spectrum}
\newacronym{dsss}{DSSS}{direct sequence spread spectrum} 
\newacronym{ew}{EW}{electronic warfare}
\newacronym{rf}{RF}{radio frequency}
\newacronym{dos}{DoS}{denial of service}
\newacronym{hitl}{HITL}{hardware-in-the-loop}
\newacronym{srn}{SRN}{Standard Radio Node}

\newacronym{mchem}{MCHEM}{massive channel emulator}

\usepackage{longtable}

\IEEEoverridecommandlockouts

\begin{document}

\title{\esword: Implementation of Wireless Jamming Attacks in a Real-World Emulated Network}

\author{\IEEEauthorblockN{Clifton Paul Robinson$^\dagger$, Leonardo Bonati$^\dagger$, Tara Van Nieuwstadt$^*$, Teddy Reiss$^*$, Pedram Johari$^\dagger$,\\Michele Polese$^\dagger$, Hieu Nguyen$^*$, Curtis Watson$^*$, Tommaso Melodia$^\dagger$}
\IEEEauthorblockA{$^\dagger$Institute for the Wireless Internet of Things, Northeastern University, Boston, MA, U.S.A.\\E-mail: \{robinson.c, l.bonati, p.johari, m.polese, melodia\}@northeastern.edu}
\IEEEauthorblockA{$^*$The MITRE Corporation, Bedford, MA, U.S.A.\\E-mail: \{tvannieuwstadt, treiss, htnguyen, cmwatson\}@mitre.org}

\vspace{-30pt}

\thanks{This work was partially supported by the U.S.\ National Science Foundation under grant CNS-1925601.
Additional funding support was provided by the MITRE Corporation. Approved for Public Release; Distribution Unlimited. Public Release Case Number 22-2965.
}
}

\maketitle

\begin{abstract} 
Jamming attacks have plagued wireless communication systems and will continue to do so going forward with technological advances.
These attacks fall under the category of Electronic Warfare (EW), a continuously growing area in both attack and defense of the electromagnetic spectrum, with one subcategory being electronic attacks (EA).
Jamming attacks fall under this specific subcategory of EW as they comprise adversarial signals that attempt to disrupt, deny, degrade, destroy, or deceive legitimate signals in the electromagnetic spectrum.
While jamming is not going away, recent research advances have started to get the upper hand against these attacks by leveraging new methods and techniques, such as machine learning.
However, testing such jamming solutions on a wide and realistic scale is a daunting task due to strict regulations on spectrum emissions.
In this paper, we introduce \esword (\textbf{e}mulation (of) \textbf{S}ignal \textbf{W}arfare \textbf{O}n \textbf{R}adio-frequency \textbf{D}evices), the first large-scale framework that allows users to safely conduct real-time and controlled jamming experiments with hardware-in-the-loop.
This is done by integrating METEOR, an \gls{ew} threat-emulating software developed by the MITRE Corporation, into the Colosseum wireless network emulator that enables large-scale experiments with up to 49 software-defined radio nodes.
We compare the performance of \esword with that of real-world jamming systems by using an over-the-air wireless testbed (considering safe measures when conducting experiments).
Our experimental results demonstrate that \esword achieves up to 98\% accuracy in following throughput, signal-to-interference-plus-noise ratio, and link status patterns when compared to real-world jamming experiments, testifying to the high accuracy of the emulated \esword setup.

\end{abstract}

\glsresetall

\begin{figure*}[ht]
    \centering
    \includegraphics[width=\textwidth]{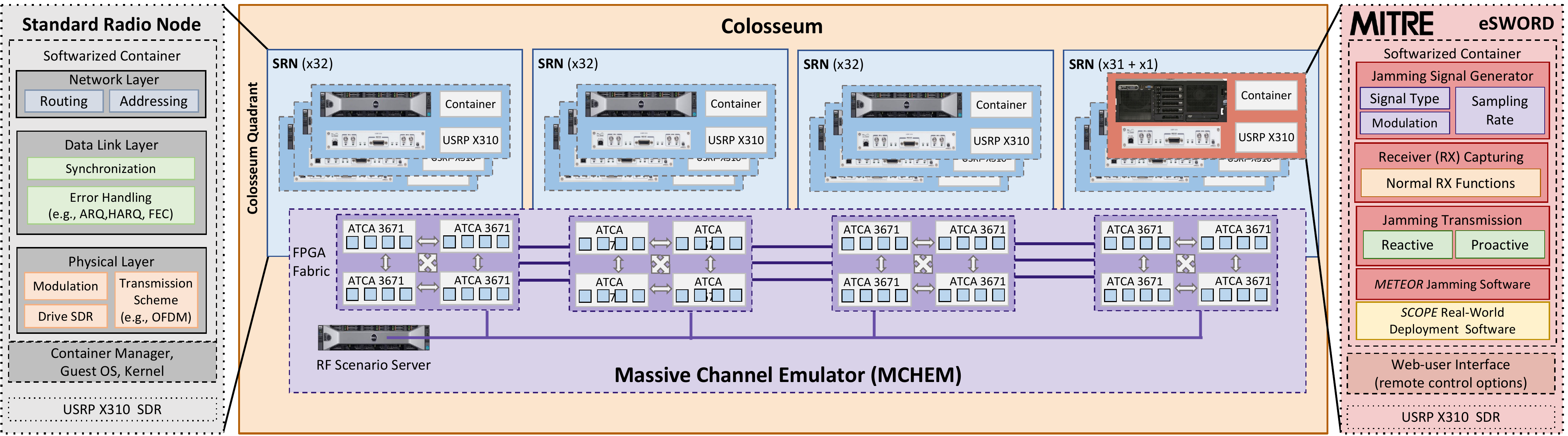}
    \caption{Overview depicting how \esword is integrated into Colosseum, taking the place of one of the SRNs, allowing for the jamming software to be used.}
    \label{fig:esword}
    \vspace{-10pt}
\end{figure*}

\section{Introduction}
\label{sec:intro}

\Gls{ew}---defined as the ability to use the electromagnetic spectrum to sense, protect, and communicate, as well as deny adversaries the means to disrupt and use these signals~\cite{ew_def}---has continuously grown throughout the years as a major area in both attack and defense~\cite{ew_def}.
The importance of \gls{ew} for both attack and defense applications is further confirmed by its steep market growth in recent years (worth \$17 billion in 2020 alone, and expected to reach \$21 billion by 2025~\cite{ew_money}).
%
\Gls{ew} is divided into three major categories, each having many subcategories: (i)~\textit{Electronic Attack}, which utilizes electromagnetic energy as an offensive weapon in combat; (ii)~\textit{Electronic Protection}, used to protect personnel and equipment from the effects of the electromagnetic spectrum; and (iii)~\textit{Electronic Support}, where the focus is on the recognition and response to these threats to help detect enemy's electromagnetic weapons.
In this paper, we primarily focus on the \textit{Electronic Attack} area. Electronic attacks are any form of adversarial signal that attempts to disrupt, deny, degrade, destroy, or deceive signals on the electromagnetic spectrum~\cite{ew_def}.
One of the most common types of these attacks comes from the broad area of signal jamming, a type of \gls{dos} attack where malicious entities block legitimate communications by causing intentional signal interference~\cite{grover_jamming_2014}.
%

Jamming attacks have plagued wireless communication systems for many years, and they continue to be a problem as technology evolves, as there is no single solution to handle this broad adversarial attack.
However, in recent years, research solutions using novel techniques applied to jamming, e.g., machine learning, have started to spawn~\cite{punal_machine_2014, upadhyaya_machine_2019, gecgel_jammer_2019, kanwar_jamsense_2021}.
Although such solutions have started to get the upper hand on jamming~\cite{arjoune_novel_2020}, prototyping and testing them at scale and in realistic wireless environments is a daunting task because of the strict regulations on spectrum emissions.
For example, the United States forbids any form of signal jamming since such signals pose a serious risk to public safety communications (e.g., they could prevent someone from making emergency calls)~\cite{noauthor_jammer_2011}.
Even though some methods exist to validate jamming solutions, e.g., software simulations or experiments in anechoic chambers, they can hardly capture either the accuracy or the scale of real networks with actual \gls{hitl}.
\gls{hitl} offers the most benefits to jamming experiments as it enables \textit{real}, non-synthetic signals over emulated channels with real hardware devices, and without causing harmful interference to public communications.
 
In this paper, we propose \esword (\textbf{e}mulation (of) \textbf{S}ignal \textbf{W}arfare \textbf{O}n \textbf{R}adio-frequency \textbf{D}evices), a first-of-its-kind, large-scale framework with \gls{hitl} to conduct real-time, accurate tests of jamming signals on a wireless spectrum. \esword allows for the testing of different adversarial jamming scenarios on a wireless spectrum with \textit{real} signals (both adversarial and legitimate) in a safe environment.
We prototype \esword on the Colosseum wireless network emulator~\cite{bonati2021colosseum}, a large-scale \gls{sdr}-based testbed that allows users to evaluate solutions in realistic but controlled environments with \gls{hitl}.
We leverage this testbed to evaluate our solution at scale over a softwarized cellular network with 50~nodes shared among base stations, users, and jammer.
Colosseum allows us to prototype \esword in a controlled environment without causing external interference to commercial devices.
We compare the results obtained with \esword to those of real wireless jamming signals (collected ethically). We verify that \esword is able to provide accurate real-time results, e.g., \esword's throughput follows a pattern with accuracy up to 98\% when compared to real-world results.
Finally, we perform a jamming experiment in a controlled over-the-air testbed in similarity to the \esword, and verify the accuracy of its results.
%
To the best of our knowledge, \esword is the first jamming emulation system that makes it possible for researchers to evaluate solutions at an accurate network scale with \gls{hitl} and in multiple emulated but realistic wireless environments provided by the Colosseum wireless network emulator.

The main contributions of this paper are as follows:
\begin{enumerate}
    \item We create and prototype \esword (Fig.~\ref{fig:esword}, explained in details in Section~\ref{sec:setup}), a framework that allows for the emulation of jamming in a wireless spectrum at scale (with up to 50 communicating nodes), where one of the nodes can be an adversarial jammer.
    \item We show that \esword is able to generate multiple types of jamming signals using different signal types and modulations to allow for many unique types of attacks.
    \item We compare \esword results with real-world jamming signals, demonstrating \esword emulation accuracy.
\end{enumerate}

The remainder of the paper is organized as follows. Section~\ref{sec:relate} discusses the related work and research regarding this area. Section~\ref{sec:jam_attack} lays out the jamming adversary we focused on in this paper. In Section~\ref{sec:problem}, we discuss the \esword prototype and the components that create it. In Section~\ref{sec:setup}, we describe the testbed implementation, while experimental results are shown in Section~\ref{sec:results}. Finally, Section~\ref{sec:end} concludes the paper.



\section{Related Work}
\label{sec:relate}

Wireless jamming has been studied for many years by the research community. There is a constant ebb and flow with research surrounding it due to the fact this form of attack advances with the advancement of the technology. While attempting to implement new methods to protect against these attacks, a major focus is discovering how these signals impact the wireless networks as a whole~\cite{benslimane_analysis_2011, wilhelm_short_2011}. Overall, wireless jamming research can be put into three broad categories, (i)~attack~\cite{xu_feasibility_2005, li_optimal_2007, wilhelm_short_2011, vadlamani_jamming_2016, proano_selective_2010, benslimane_analysis_2011, raymond_denial--service_2008}; (ii)~defense~\cite{grover_jamming_2014, li_optimal_2007, pirayesh_jamming_2022}; and (iii)~detection~\cite{xu_feasibility_2005, cakiroglu_jamming_2008, aschenbruck_simulative_2010}.

Jamming attacks and defenses usually go hand-in-hand within the research community as defense techniques, such as signal detection and frequency hopping~\cite{grover_jamming_2014, pirayesh_jamming_2022}, are investigated to mitigate the considered attacks~\cite{grover_jamming_2014, li_optimal_2007, pirayesh_jamming_2022}. In these experiments, there is a true focus on ``knowing your enemy'' by taking a deep dive into these sorts of attacks and seeing how they work within a network \cite{xu_feasibility_2005, wilhelm_short_2011, benslimane_analysis_2011}.
Large-scale WiFi research has been done in the past, showing how jamming attacks on commercial wireless solutions, such as those enabled by the IEEE 802.11 standard, deteriorate the network performance~\cite{benslimane_analysis_2011}, and showing that high WiFi data rates are not resilient to jammers.
Further research focuses on specific types of jamming implementations.
The authors of~\cite{proano_selective_2010} study selective jamming, where the adversary focuses on ``high-value'' targets by exploiting their knowledge of the network, while also proposing a prevention mechanism that neutralizes the inside knowledge of the attacker.
Still a common jamming technique today, reactive jamming is instead known for its strategy and detection avoidance. For instance, the authors of~\cite{wilhelm_short_2011} discuss the implications of this form of attack in wireless networks.

Jamming avoidance has become more advanced with the use of spread-spectrum techniques~\cite{grover_jamming_2014, pirayesh_jamming_2022, zou_survey_2016}. The two main techniques used today are \gls{fhss} and \gls{dsss}. The former allows for jamming avoidance as it can literally ``jump'' away from attacks by hopping to frequencies not affected by jammers~\cite{freq_hop}. The latter takes a different approach, using rapid phase transitions with the data, spreading it on a larger bandwidth, thus conferring it more resilience to jammers~\cite{dsss}.


While extensive research surrounding jamming has been done throughout the years, to the best of our knowledge, existing implementations and techniques are tested either through software simulations or in small-scale setups.
%
In this sense, our research takes a step forward by implementing and evaluating \esword on a large-scale testbed with hardware-in-the-loop.
As opposed to software-based simulations, this gives us access to data inputs from real physical devices, i.e., \glspl{sdr}, and allows for experiments in controlled, but realistic environments without compromising commercial systems.


\section{The Adversary: Wireless Jamming}
\label{sec:jam_attack}

Jamming attacks have been exacerbated in recent years as the technology needed to create jamming signals has become more accessible and affordable~\cite{grover_jamming_2014}.
Today, anyone with a \gls{sdr} and a few lines of code can create a jammer that can deteriorate the performance of a wireless network, or even annihilate it.
On a much larger scale, jamming has become a popular form of \gls{ew}~\cite{thurbon_origins_1977}. Such attacks can be coordinated better, with novel implementations that can take down large portions of cellular, GPS, and wireless networks in a deployed area.

\begin{figure}[t]
    \centering
    \includegraphics[width=\columnwidth]{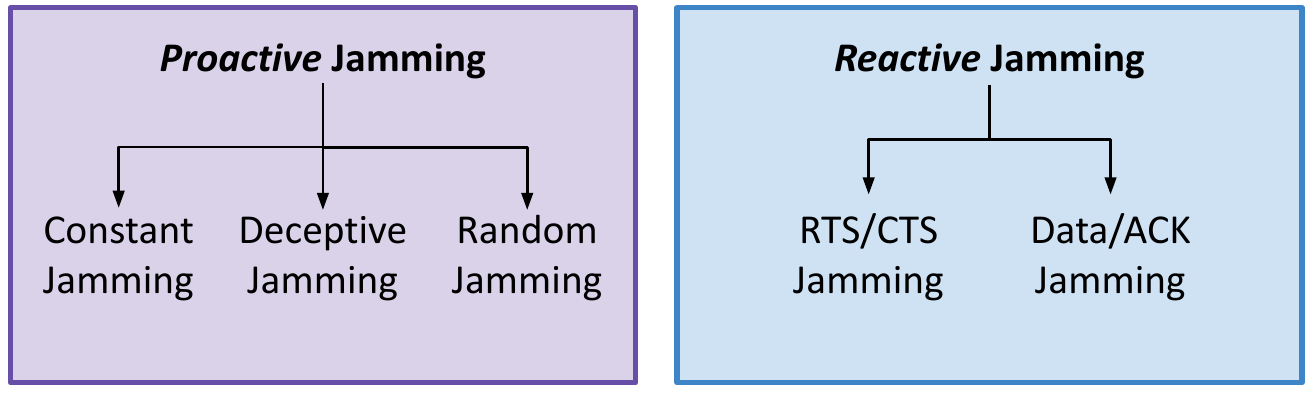}
    \vspace{-15pt}
    \caption{Two broad categories of different jamming attacks.}
    \label{fig:jam_types}
    \vspace{-15pt}
\end{figure}

Fig.~\ref{fig:jam_types} shows two of the most common categories of jamming attacks which is our focus in this paper. Proactive jamming is the most simple of the two, it happens when a signal is transmitting whether there is communication in a network or not~\cite{grover_jamming_2014}. Reactive jamming goes a step further, as jamming signals are only transmitted when communications are occurring in the network~\cite{wilhelm_short_2011}. This makes it more difficult to detect adversaries as reactive jamming waits for the presence of data. 

A proactive jammer works by placing the transmitters in a specific channel into non-operating mode, making that channel unusable~\cite{grover_jamming_2014}.
A reactive jammer, instead, actively monitors a predetermined bandwidth and responds to specific packets or signals that are currently traveling in the air~\cite{wilhelm_short_2011}.
Unlike its proactive counterpart, a reactive jammer can be harder to detect as it targets on-air packets and can tune the time and spectrum location of the attack.
Reactive jamming is also considered a stepping-stone, as it is the most common form of jamming that is used to implement more optimal jamming attacks~\cite{wilhelm_short_2011}.

An example of our reactive jammer on a 10MHz channel can be seen in Figure \ref{fig:jam_signal}. In this example, a jamming signal with a bandwidth of 156KHz starts attacking the WiFi signal on the center frequency 2.378GHz, and after $\sim$100\textit{ms} the WiFi signal shifts to the center frequency 2.382GHz, where the jammer shortly follows after it, sensing the change in energy location. This attack has direct impacts on the WiFi signal's transmissions, as when it avoids the signal has throughput values up to 11Mbps, but when the jammer catches it can drop all the way to 4Mbps.
By using both proactive and reactive attacks, we cover a broad scope of knowledge within the attack area.
%


\section{Our Prototype: \textit{eSWORD}}
\label{sec:problem}

\esword is a software prototype that utilizes jamming software
within a large-scale network emulator. The main goal of \esword is to provide a means to run large-scale jamming experiments in an accurate and safe environment.

Fig.~\ref{fig:esword} shows the high-level overview of our prototype (``\esword Device'' in the figure). The prototype comes with a proprietary jamming signal generator, where the signal type, modulation, and sampling rate can be easily adjusted. Normal TX/RX functions are also included to transmit data, as well as jamming transmission capabilities for multiple forms of attacks. By utilizing this jamming software, we can create adversarial signals that work against practical real-world systems. As stated in Section \ref{sec:jam_attack}, our threat emulator focuses on proactive and reactive attacks. This allows for the implemented jammer on the emulator to give broad results that cover a multitude of similar attacks. As this threat emulator is designed to work in the real-world, commercial \gls{sdr} hardware enables the use of \esword over-the-air on real wireless networks, granted the appropriate steps are taken to ensure safe transmissions.

\begin{figure}[!h]
    \centering
    \vspace{-5pt}
    \includegraphics[width=\columnwidth]{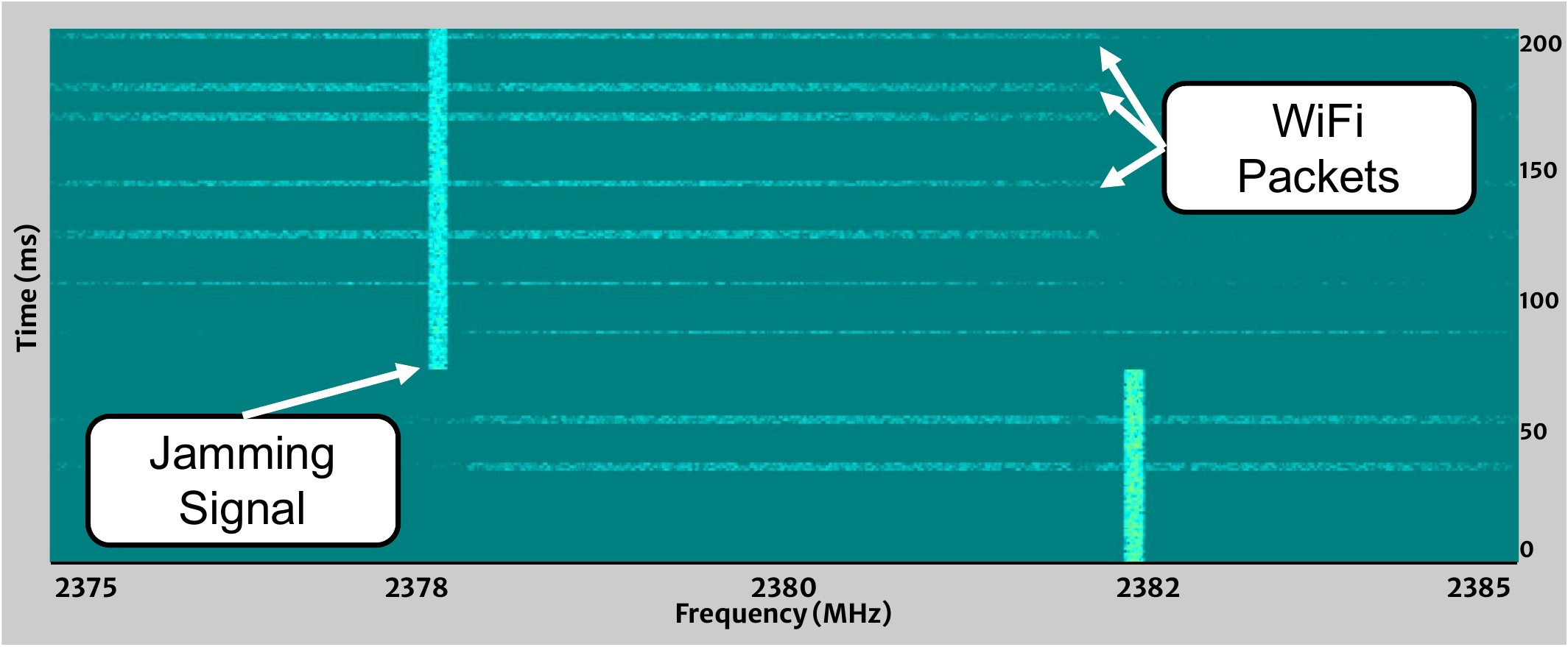}
    \caption{An example of reactive jamming on frequency hopping WiFi signals.}
    \vspace{-8pt}
    \label{fig:jam_signal}
\end{figure}

Our prototype is controlled through a web interface that supports RESTful APIs and gives control of the \gls{sdr} to the users. By using such an interface, users can control when data streams begin and end, as well as when jamming signals are transmitted. Within both the RX and TX, the center frequency, sampling rate, and gain can all be adjusted. In addition, a live spectrogram (see Fig.~\ref{fig:jam_signal}) can be started together with the RX stream to easily monitor the received signals in real-time.

\subsection{Jamming Software \& Attack Vector}
\label{sec:attack_vector}

Our prototype utilizes an \gls{ew} threat-emulating software, called METEOR, developed by the MITRE Corporation~\cite{mitre}.
METEOR is built with commercial off-the-shelf \gls{sdr}, supporting software-defined capabilities and programmability.
\esword uses the \gls{ew} attack capabilities of METEOR, which include reactive and proactive jamming attacks in wireless scenarios.
Such an \gls{ew} attack could indeed be used in the wild to hamper or disrupt complex wireless communication systems.

\begin{table}[!b]
\setlength\belowcaptionskip{5pt}
    \centering
    \footnotesize
    \setlength{\tabcolsep}{2pt}
    \caption{Types of jamming signals and modulations supported by METEOR.}
    \label{tab:jam_sigs}
    \begin{tabularx}{\columnwidth}{
        >{\raggedright\arraybackslash\hsize=0.8\hsize}X 
        >{\raggedright\arraybackslash\hsize=1.2\hsize}X }
        \toprule
        Signal Type & Modulation \\
        \midrule
        FSK & Frequency shift keying \\ 
        ASK & Amplitude shift keying \\
        MSK & Minimal shift keying \\
        AWGN & White noise (barrage jamming) \\
        Band Noise & Band-limited noise signal \\
        Continuous-Phase & Continuous-phase frequency shift keying \\
        \bottomrule
    \end{tabularx}
\end{table}

METEOR offers a versatile jamming file generator that supports diverse types of modulated signals, shown in Table~\ref{tab:jam_sigs}.
%
%
These different types of signals allow for multiple forms of jamming attacks to be created, including but not limited to constant jamming signals, pulsed signals, and narrowband or wideband signals. The specific signals we leverage for the attacks carried out in this paper are narrowband jamming signals with proactive and reactive capabilities.
This variety of modulation types creates a large attack vector that allows for testing the resilience of different kinds of networks against specific attacks.
On top of its signal generation capabilities, METEOR also allows users to upload their own custom files to use as jamming signals.


%

\section{Experimental Testbed Setup}
\label{sec:setup}

We implement \esword on the Colosseum wireless network emulator~\cite{bonati2021colosseum}, a publicly available testbed that allows users to perform large-scale experiments with up to 128~programmable nodes and radio devices (see Fig.~\ref{fig:esword}).
%
Specifically, Colosseum is formed of two main blocks: (i)~the \glspl{srn}, and (ii) the \gls{mchem}.
The \glspl{srn} (shown in Fig.~\ref{fig:esword}, top-middle) are 128~remotely accessible compute nodes that control a USRP~X310 \gls{sdr} each.
These nodes can be leveraged to perform custom experiments through softwarized protocol stacks (Fig.~\ref{fig:esword}, left) executed on them and used to control the \glspl{sdr} that act as radio front-ends.
\gls{mchem} (Fig.~\ref{fig:esword}, bottom-middle), instead, takes care of emulating channel conditions between every pair of \glspl{srn} by processing signals generated by the \glspl{sdr} through FPGA-based \gls{fir} filters.
These FIR filters apply the channel impulse response of the Colosseum scenario selected by users to the \gls{sdr} signals, thus emulating the actual real-world wireless scenario.
We use METEOR (Fig.~\ref{fig:esword}, right) to interfere with a network composed of 49~nodes implemented through the SCOPE framework~\cite{bonati_scope_2021}.
This framework extends srsRAN---which allows users to deploy cellular protocol stacks on software-defined nodes and radios---with automated pipelines to swiftly run on the Colosseum testbed.
Colosseum enables extensive testing environments and conditions through a set of diverse wireless scenarios
representative of real-world urban cellular deployments, channels, and traffic~demand.

%
To perform our jamming experiments, we integrated the METEOR node and dedicated \gls{sdr} (a USRP~X310) within Colosseum, which enables us to potentially jam any of the Colosseum nodes. 
In our experiments, we considered center frequencies of 980 MHz for the cellular uplink signals and 1020 MHz for the downlink signals.
Fig.~\ref{fig:esword} gives an overview of our \esword prototype framework that integrates the METEOR jamming system into the Colosseum wireless network emulator.
For the jammer to be used, the jamming software is flashed onto the FPGA of \esword \gls{sdr}, and driven by a dedicated compute node.
With METEOR being integrated to the channel emulator, it can operate similarly to the normal nodes on the Colosseum, except instead of being an \gls{srn}, it represents a jammer node that can jam the signals traveling across the scenario emulated by Colosseum. 
As of now, \esword includes a single jammer; however, multiple jammers can be integrated following a similar procedure if needed.

\subsection{Node Placement within Testbed}\label{sec:node-cluster}

\begin{figure}[!b]
    \centering
    \vspace{-5pt}
    \includegraphics[width=\columnwidth]{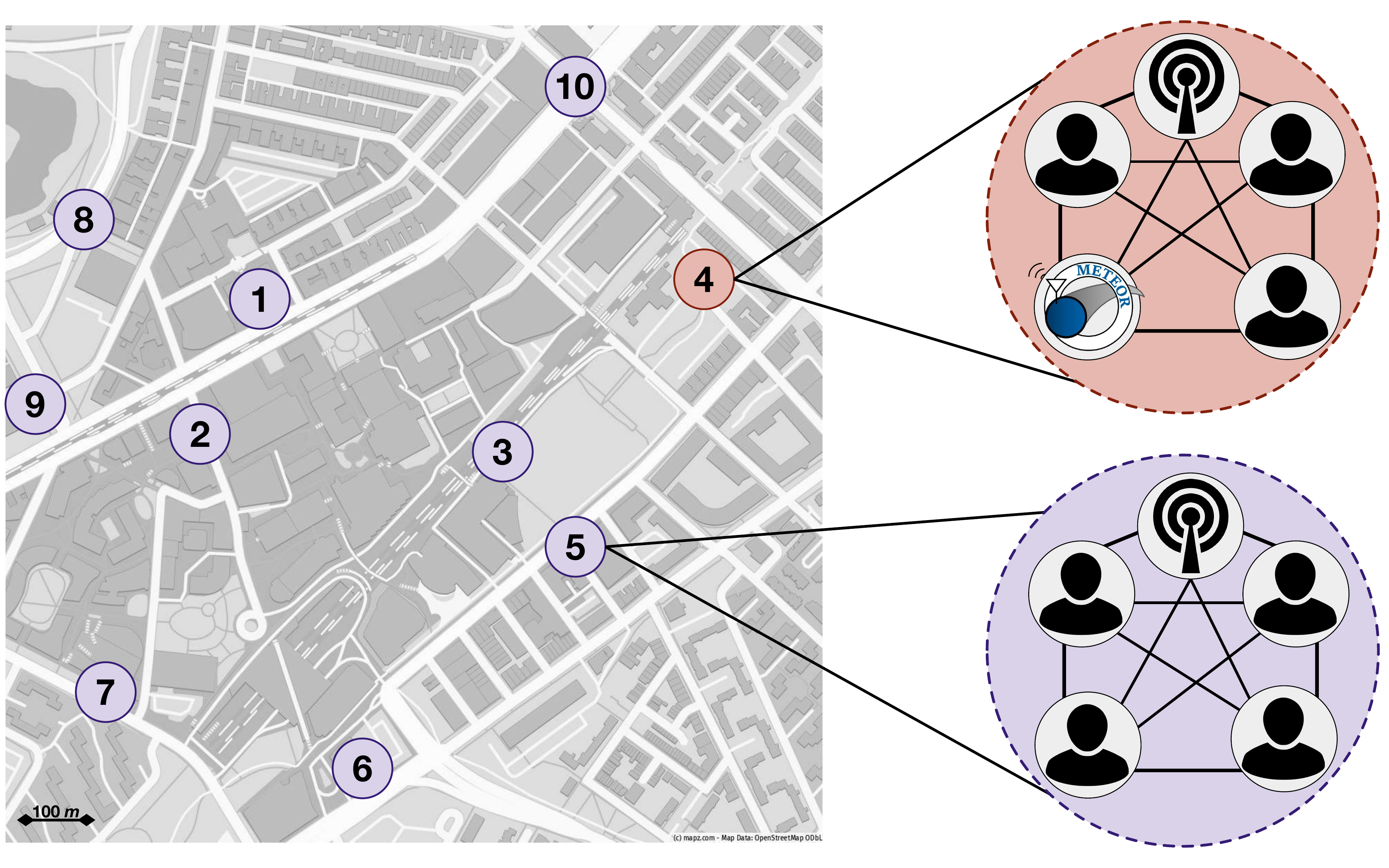}
    \vspace{-15pt}
    \caption{The real-world example of node placement in Boston, with a jamming node located in cluster \#4 taking a spot out of the 50 possible nodes.}
    \label{scope}
\end{figure}

To emulate a diverse set of wireless environments, we leveraged the large-scale urban cellular scenarios available on Colosseum~\cite{bonati_scope_2021}.
These scenarios allow for real-world location testing, using base stations whose locations match the coordinates of commercial deployments, as well as a number of user equipment deployed in their surroundings.
These locations correspond to Rome, Italy; Boston, U.S.; and Salt Lake City, U.S. A sample map of the locations of the base stations considered for the Boston area is shown in Fig.~\ref{scope}.
This scenario includes 10~cellular base stations, and 40~cellular users (deployed in groups of 5, i.e., 1 base stations around 4 users in each cluster).
Base stations and the users they are serving divide the network into clusters (see Fig.~\ref{scope}).
By using these scenarios, we can test jamming impacts across different base stations, as well as move the jamming node to different locations within the scenario.
%
In our experiments, a single node is replaced with a malicious node with jamming capabilities. This setup is able to show how a single jamming source impacts the initial user placement, as well as the base stations around it.

\subsection{Over-The-Air Experimental Setup}

We leverage the Arena testbed~\cite{arena} to perform our over-the-air experiments. Arena consists of a grid of USRP X310 \glspl{sdr} deployed in an indoor office environment.
To conduct our real-time, non-emulated experiments, we use 3~of Arena \glspl{sdr} and perform safe narrowband jamming experiments that do not interfere with any of the spectrum used by external applications.
One of Arena \glspl{sdr} acts as a receiver, one as a transmitter, and one as a jammer. The jamming signal used in this case is built to mimic the characteristics of the signals of the \esword Colosseum prototype.
In this case, we collect channel throughput statistics and compare to those of the \esword.


\section{Experimental Results}
\label{sec:results}

The results collected in these experiments are designed to test the accuracy of \esword. The first part of these results compares the \esword prototype implemented on Colosseum with real-world jamming signals, while the second part discusses the impact on nodes in the emulated scenario.
In both cases, \esword is used to jam a cellular network with 49~software-defined nodes (10~base stations and 39~users, see Section~\ref{sec:setup}).

\vspace{-8pt}

\subsection{Over-the-Air vs. Emulated Environment}

In this section, we validate the accuracy of \esword prototype on the Colosseum emulator, comparing it with real-world---over-the-air---jamming signals, used as our baseline.

\subsubsection{Jamming Signal Composition}
The jamming signal for both real-world and emulation setups are constructed in the same manner (i.e., center frequency, sampling rate, etc.), allowing for accurate emulation, with the option of adjusting sampling rate, gain, and signal size. In this specific example, both signals are narrowband noise jammers using FSK modulation, focusing on disrupting a single channel's communication.

\begin{figure}[!t]
    \centering
    \includegraphics[width=.98\linewidth]{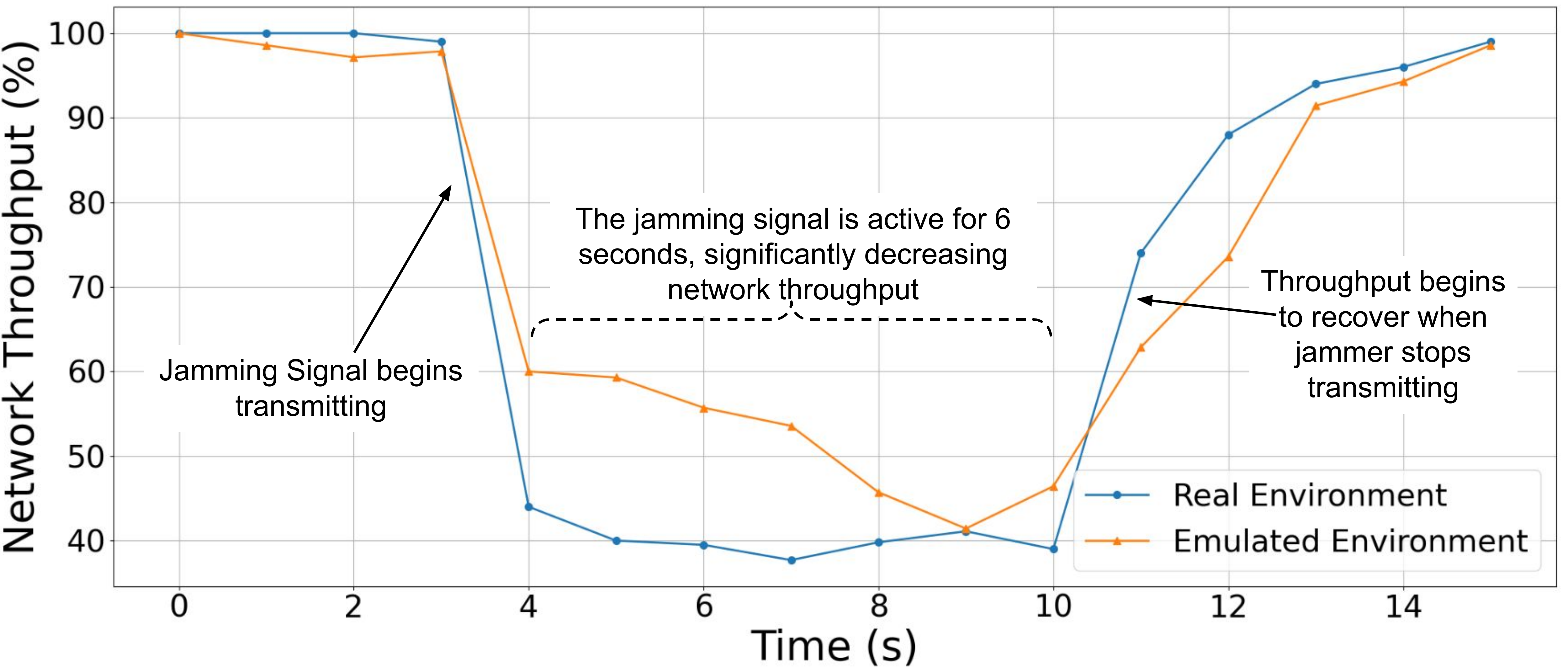}
    \caption{A throughput example of a real-world network impacted by a jamming signal and the \textit{eSWORD} emulated network impacted by the jamming signal. 100\% refers to the throughput of the legitimate nodes in absence of the jammer.}
    \label{tp}
    \vspace{-10pt}
\end{figure}

\subsubsection{Channel Throughput Comparison}
Throughput is one of the main metrics used to test the performance of a network.
Fig.~\ref{tp} shows the throughput (expressed as a percentage) in both the real and emulated networks over a 15-second window when a jammer is introduced. This jammer is stationary and holds the same power level while turned on.
%
It can be seen that for both networks, the throughput stays near the maximum value (i.e., 100\%) for the first 3 seconds of the experiments, until a narrowband jamming signal is introduced between seconds 3 and 4.
Once the jammer is activated, the throughput significantly drops for both cases.
The real network environment drops by almost 60\% as the jammer interrupts the ongoing communications, while the drop in the emulated environment is slower, which can be attributed to the node locations differing in closeness, with the throughput decreasing by 40\% within the first second.
As the constant jamming signal continues, both networks drop to similar points (i.e., both having throughput drops near 60\% of the original signal). Once the jammer is deactivated (second 10 of the experiment), both networks quickly recover back to their original throughput values. For the accuracy of the throughput between the two environments, the comparative accuracy of the emulated data is between 75\% and 98\% as shown in Fig.~\ref{tp}.

\subsection{Impact on Radio Communication Performance}
\begin{figure}[!b]
    \vspace{-5pt}
    \centering
    \includegraphics[width=.98\columnwidth]{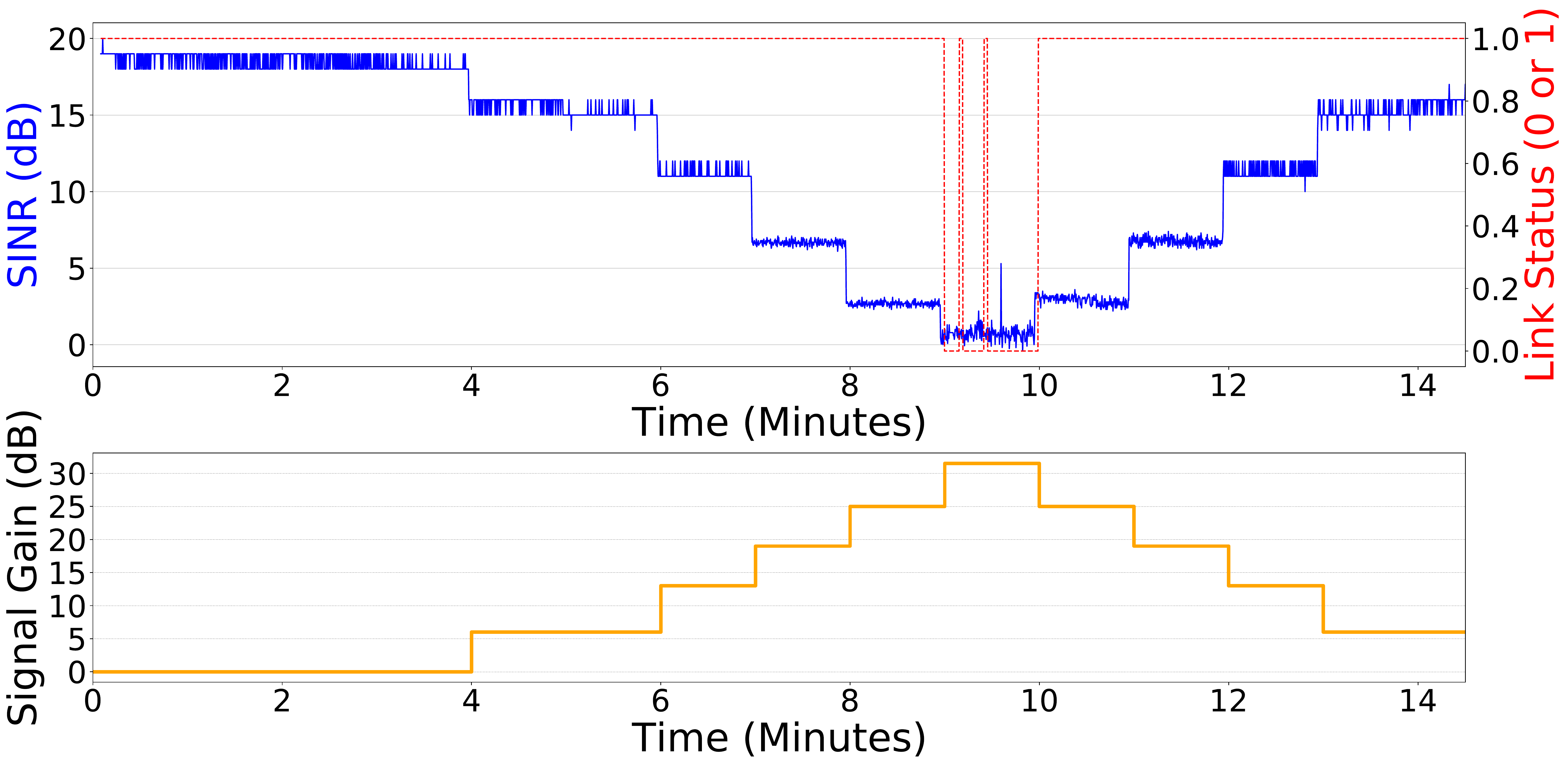}
    \caption{Top: impact a jammer has on a node's link status and SINR based on the gain of the signal over time. Bottom: the jammer's change in gain over time, showing how the SINR of a node and the signal gain mirror each other.}
    \label{fig:gain}
\end{figure}

Emulated environments require more than just an accurate-looking jamming signal to claim they are correctly functioning.
Indeed, there also needs to be actual cause and effect to the network and nodes as well (i.e., the way a jammer is set up impacts the network in different ways).
%
As an example, the gain of the jamming signal has a direct impact on how a legitimate node is affected, as lower gains impact less than higher ones.
This also determines whether a node will still be able to transmit while being affected by the jamming interference.
Fig.~\ref{fig:gain} shows the real-time impact of different gains on a legitimate node.
We consider how a jamming signal varying its gain from 0 to 32 dB (shown at the bottom of the figure) impacts the \gls{sinr} of a legitimate node, and its link status, in a time window of 15 minutes (top portion of the figure).
We notice that the \gls{sinr} gradually decreases toward 0 with each increase in the jamming gain, showing the direct impact of the jammer on the legitimate node.
Specifically, when the gain of the jammer is highest (between minutes 9 and 10), the link status of the legitimate node drops to 0, effectively causing the node to detach from the network (and not being able to communicate with the remaining nodes).
%
As the jamming gain lowers, the \gls{sinr} of the node improves, and the node is able to reconnect to the network.
%
%
From this experiment, we notice that as the gain of the jammer increases by 5~dB at each time-step, the \gls{sinr} of the legitimate node drops between 16\% to 20\% which shows an inverse correlation between the jamming power and the achievable \gls{sinr}.

\subsection{Impact on Node Clusters}
The node clustering (discussed previously in Sec.~\ref{sec:node-cluster}, and shown in Fig.~\ref{scope}) gives a perspective on the real-world impact a jammer can have on different node clusters.
In the real world, wireless jammers only impact the areas in which they are deployed. 
However, depending on the signal strength, surrounding areas can be affected as well.
In this experiment, we deploy \esword in the Colosseum scenario (shown in Fig.~\ref{scope}) and evaluate how the jammer works in the emulated environment.

\begin{figure}[!t]
    \centering
    \includegraphics[width=\columnwidth]{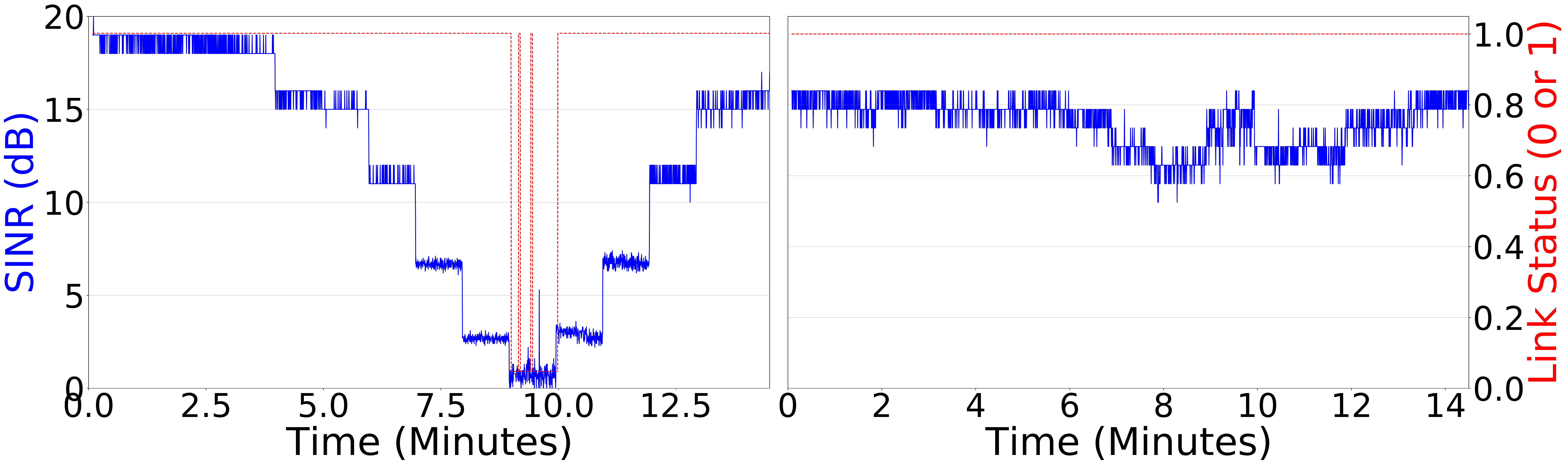}
    \caption{Two nodes where one (left) is in the same group as the jammer and the other node (right) is relatively close, but not in the same group.}
    \label{node_impact}
    \vspace{-10pt}
\end{figure}

Fig.~\ref{node_impact} shows the \gls{sinr} and link status for two nodes belonging to different clusters of the network.
The node on the left of the figure belongs to the same group as the jammer; the one on the right is close to the jammer but does not belong to the same cluster.
Looking at the node in the same cluster of the jammer (left), we notice that the jammer impact is almost immediate.
Indeed, once the jammer is started, the legitimate node can no longer communicate with the other nodes of the network.
We also note that the \gls{sinr} of this node assumes values lower than 0 at times, meaning that the strength of the jamming signal is stronger than that of the legitimate signals received.

Conversely, the node in the different cluster (on the right in the figure) experiences a less severe signal degradation and can communicate throughout the experiment.
This is due to the fact that even though we notice several instants in which the \gls{sinr} of the node decreases (e.g., at minutes 7 to 9, and 14-15), meaning that the jammer can reach nodes further away, it does so with limited signal strength that does not interrupt the ongoing communications.
%
Overall, we observe that the node in the different cluster (which is further away from it) only experiences a \gls{sinr} degradation up to 38\%, while the node in the same cluster (closer to the jammer) experiences a \gls{sinr} drop up to 70\%, and in multiple instants of the experiment.

These experiments demonstrate that the \esword prototype implemented on the Colosseum wireless network emulator allows us to perform realistic experiments in a controlled environment.
%
Concerning the location of the critical infrastructure, \esword allows users to not only jam nodes in close proximity to the jammer, but also those further away. This reflects on the deployment of real-world network nodes and components, which may not be confined to a single location, but may be across different ones.
Adversaries can exploit this knowledge to aim to take out specific portions of the infrastructure, and potentially cripple the entire network.
By enabling the testing of jamming attacks in controlled but realistic environments, \esword allows users to evaluate the resilience of such critical network systems without harming commercial infrastructures, and to find robust ways to counter such forms of attack.




\section{Conclusion}
\label{sec:end}

In this paper, we introduced \esword, a novel framework that allows for accurate, large-scale jamming attack experimentation with HITL in a controlled environment.
We prototyped \esword on the Colosseum wireless network emulator and demonstrated its capabilities on a large-scale network with 49 cellular nodes deployed on realistic urban wireless scenarios.
Finally, we verified the accuracy of \esword results by comparing them with those obtained in an over-the-air wireless testbed.
\esword's capabilities, scale, and controlled experimentation are key in advancing jamming research, for instance, to devise techniques to counter jamming attacks and to evaluate network resilience to them, which future works will focus on.


\balance
\footnotesize
\bibliographystyle{IEEEtran}
\bibliography{mybib} 
\end{document}